\documentclass[twocolumn,showpacs,aps,prl,superscriptaddress]{revtex4}
\usepackage{graphicx} 
\usepackage{dcolumn}
\usepackage{epsfig}    
\usepackage{amsmath}

\input babarsym
\newcommand{\etal}{\emph{et al.},}
\newcommand{\calB}{\mbox{${\cal B}$}}

\begin{document}
\begin{flushleft}
~\\
~\\
~\\
\end{flushleft}

\begin{flushright}
\babar-PROC-10/21 \\
\end{flushright}

\title{Recent Results in Charmonium Spectroscopy at $B$-factories} 

\author{Pietro Biassoni \\
 (From the $\babar$
  Collaboration)}
\affiliation{INFN and Dipartimento di Fisica, Universit\`a di
  Milano, I-20133 Milano, Italy }

\begin{abstract}
\begin{center}
\emph{Contributed to the Proceedings of Flavor Physics and
\CP\ Violation Conference 2010, \\
25--29 May 2010, Turin, Italy}
\end{center}
  Charmonium spectroscopy gained renewed interest after the
  discovery in 2003 of the unpredicted $X(3872)$ charmonium-like state,
  above the $D\overline{D}$ threshold. 
  To date many charmonium-like states above the $D\overline{D}$
  threshold have been claimed. Some of these states do not easily fit
  the conventional charmonium picture. In this article we review recent 
  experimental results in this field, reported by the \babar\ and Belle
  Collaborations.
\end{abstract}

\pacs{13.25.Gv,14.40.Pq}
\maketitle

\section{Introduction}
Since the discovery of the \jpsi\ in 1974~\cite{jpsi}, the study of
charmonium spectroscopy has provided crucial tests of the quark model and
of models of the QCD interaction. The charmonium spectrum cannot be derived
directly from the QCD Lagrangian, due to the presence of large
non-perturbative effects at this energy. The charmonium spectrum may be
predicted by constituent quark model inspired potential
models~\cite{qwg} that include a Coulomb-like 
one-gluon-exchange term at short distance and a linear confining
contribution in the long range. In this picture each state is
characterized by $J^{PC}$ quantum numbers, where $J = L + S$, $P =
(-1)^{L+1}$, $C = (-1)^{L+S}$, $L$ is the orbital angular momentum,
and $S$ the total spin. Allowed $J^{PC}$ quantum numbers in
constituent quark model are $0^{\pm+}$, $1^{\pm -} $, $1^{++}$,
$2^{++}$, ... . Charmonium resonances are expected to be narrow below
the $D\overline{D}$ threshold, and wide above, due to  the accessibility
of Zweig favored decays.

In 2003, the Belle Collaboration reported the observation of a narrow
charmonium-like state decaying into $\jpsi\pip\pim$, with a
mass of about 
$3872$~\mevcc~\cite{BelleX}. This state, dubbed $X(3872)$, was then
confirmed by the 
CDF~\cite{CDFX}, D0~\cite{D0X}, and \babar~\cite{BabarX} Collaborations.
This state is above the open-flavor threshold, but has
a width of only  $3.0^{+2.1}_{-1.7}$~\mevcc~\cite{BabarXwidth},
and cannot easily fit into the 
charmonium picture provided by the potential model. The search for the
same final state in initial state radiation (ISR) production resulted
in the observation by \babar\ of the $Y(4260)$ with a mass of about
$4260$~\mevcc\ and a width of about 90~\mevcc~\cite{BabarY}. This
state has been confirmed by the CLEO~\cite{CLEOY} and Belle~\cite{BelleY}
Collaborations. Other states decaying into $\jpsi\pip\pim$
and $\psi(2S)\pip\pim$ have been claimed or observed in ISR
production~\cite{BelleY,OtherY}. 

Several new states have been observed by both \babar\ and Belle in the
3940~\mevcc\ region, including $X(3915)$~\cite{Belle3915}, $X(3940)$,
$X(4160)$~\cite{BelleDC}, $Y(3940)$~\cite{Belle3940,Babar3940}, and
$Z(3930)$~\cite{BelleZ,BabarZ}.  
A narrow resonance-like state decaying into $\jpsi\phi$ and dubbed
$Y(4140)$ was claimed by CDF~\cite{CDFPsiPhi}, but not confirmed by
Belle~\cite{BellePsiPhi,BellePsiPhigg}. However, Belle reports
evidence of another resonance, the $X(4350)$, in the same final
state~\cite{BellePsiPhigg}.  
In 2007, the Belle Collaboration claimed the observation of a narrow
state carrying hidden charm and decaying to an electrically charged final
state~\cite{BelleZch}. This state, dubbed $Z(4430)$, was 
searched and not confirmed by \babar~\cite{BabarZch}. Two more charged
charmonium-like states were reported by Belle~\cite{BelleZchich}.

The theoretical interpretation of these new states is not trivial and
often not widely accepted by the whole scientific community. There
are some 
exotic states that are predicted by QCD, but were never clearly
established. The presence of the gluon field in the QCD Lagrangian
allows for the existence of mesons with an excited gluonic degree of
freedom (hybrids)~\cite{hybrids}. The possible existence of
multi-quark states, including tetraquarks~\cite{polosa,tetraquarks},
pentaquarks~\cite{pentaquarks}, hexaquarks~\cite{exa}, and loosely
bound $D\overline{D}$ 
molecules~\cite{molecules,page,swanson} have been  investigated. Recently the
possible existence of \emph{hadrocharmonium} states, i.e. conventional
\ccbar\ states coated by light hadrons, has been 
studied~\cite{hadrocharmonium}. Finally, these states may originate
from threshold rescattering or channel coupling~\cite{threshold}.
The unusual properties of the new
charmonium-like states, such as small width and large branching ratios
into final states without open charm, may be explained inside these
exotic frameworks. It should be noted that
only multi-quark states and \emph{hadrocharmonium} are compatible with
non-zero electric charge.  

In this article, I will review recent experimental results about
new charmonium-like states. Conventional charmonium spectroscopy is
not covered. Furthermore, due to space limit, I will not cover results
related to $X(3940)$, $X(4160)$, and the $J^{PC} = 1^{--}$ states
observed in the 4260--4660~\mevcc\ region. 
A comprehensive review of charmonium spectroscopy may be found
in~\cite{polosa,eichten}.

\section{Experimental techniques}
Results presented in this paper are obtained using data collected by
\babar, Belle and CDF detectors, whose detailed description may be
found in~\cite{babarNIM,belleNIM,cdfNIM}, respectively.
At CDF a proton and an anti-proton are collided at a center of mass
energy $\sqrt{s}=1.96$ TeV. The \ccbar\ states are produced both
promptly or in $B$ meson decays. At $B$-factories, several production 
processes may be exploited:
\begin{itemize}
\item $B$ meson decay: exotic \ccbar\ states are studied in
  $B\to(\ccbar)K^{(*)}$ with 
  the subsequent decay of \ccbar\ into the desired final state. The $B$
  meson is discriminated against \qqbar\ background using kinematic
  variables such as $\Delta E\equiv E_B-\frac{1}{2}\sqrt{s}$ and $\mes
  \equiv \sqrt{s/4 - \vec{p}_B^2}$, where $(E_B,\vec{p}_B)$ is the $B$
  meson four-momentum vector expressed in \FourS\ rest frame and
  $\sqrt{s}$ is the center-of-mass energy;
\item ISR production: \ccbar\ is produced in
  $\epem\to\ccbar\,\gamma_{ISR}$, where $\gamma_{ISR}$ is a photon
  radiated by the incoming electron or positron. Since \ccbar\ couples
  directly to the virtual photon, $J^{PC}=1^{--}$.
\item Two-photon collision: \ccbar\ is produced in the
  $\epem\to\gamma\gamma\epem\to\ccbar\epem$ process. Usually the outgoing
  electrons are not detected, so the photons are quasi-real. This
  implies that the accessible $J^{PC}$ numbers are $0^{\pm+}$ and
  $2^{\pm+}$~\cite{yang}. Spin-parity numbers with $J>2$, that are
  allowed in two-photon production, are expected to be suppressed, due
  to the limited phase space available to the decay.
\item Double charmonium production: the \ccbar\ state is produced
  recoiling against a \jpsi\ in the $\epem\to\jpsi(\ccbar)$
  process. $C$-parity is positive. So far only $J=0$ states have been
  observed.    
\end{itemize}

The \ccbar\ decay is usually fully reconstructed. Inclusive searches
have been performed in double charmonium production~\cite{BelleDC}.
Information on $J^{PC}$ numbers may be inferred from the properties of the
production mechanism and the studied final state, and by studying the
angular distribution of \ccbar\ decay products.

\section{$\boldmath{X(3872)}$}
Quite a large number of experimental results is available for the
$X(3872)$. At $B$-factories, $X(3872)$ has been searched in $B$ meson
decays. The $\jpsi\pip\pim$ decay mode, where this state was
discovered~\cite{BelleX}, also provides insight on its nature. 
The $\pim\pim$ mass distribution is consistent with subthreshold
$\rho$ production~\cite{CDFrho}.
The CDF Collaboration has performed a full angular analysis of
the decay products, excluding all the possible $J^{P}$ numbers, but
$1^+$ and $2^-$~\cite{CDFang}. Belle has performed a spin-parity
analysis using $256~\invfb$~\cite{BelleXang} of
data. The number of 
reconstructed $X(3872)$ is insufficient to perform a full angular
analysis. Instead, a study of angular distributions suggested
in~\cite{rosner} favors a $1^{++}$ assignment. It should be noted
that the angular analysis performed is insensitive to $J >
1$. However, the study of $\pip\pim$ mass distribution gives some
insight on the state parity. In fact, the mass distribution near the
kinematic endpoint is suppressed by a centrifugal barrier
factor. Belle analysis agrees with $P=+1$ at 28\% level (agreement
with $P=-1$ is at 0.1\% level). This excludes $J^{PC}=2^{-+}$. 

Decay to $D^0\overline{D}^0\piz$ has been observed~\cite{BelleXddp} and
interpreted as $D^0\overline{D}^{*0}$. This interpretation has been confirmed in
subsequent analyses~\cite{Xtoddst}.

Belle reported evidence of the decay into
$\jpsi\pip\pim\piz$~\cite{BelleXomega}. The three pion mass
distribution is consistent with subthreshold $\omega$ production. 
The ratio of branching fractions 
$\calB(X(3872)\to\jpsi\pip\pim\piz)/\calB(X(3872)\to\jpsi\pim\pim)$ is
$1.0\pm0.4(stat)\pm0.3(syst)$. Recently, \babar\ has reported 
evidence of decay into the $\jpsi\omega$ final state with a 
significance of 4 standard deviations ($\sigma$)~\cite{BabarXomega}. The
angular distribution of the three-pion system strongly supports their
origin from an $\omega$. The ratio
$\calB(X(3872)\to\jpsi\omega)/\calB(X(3872)\to\jpsi\pim\pim)$ is equal to
$0.7\pm0.3$ and $1.7\pm1.3$, for charged and neutral $B$ decays,
respectively, where the error is the sum in quadrature
of the statistical and
systematic uncertainties. An analysis of the three pions mass
distribution, similar to 
the one in~\cite{BelleXang}, is performed. The agreement with $P=-1$
(61.9\%) is far better than with $P=+1$ (7.1\%). Thus, $J^{PC}=2^{-+}$
is favored, in contrast with Belle~\cite{BelleXang}.

Radiative $X(3872)$ decay into $\jpsi\gamma$ has been observed by
\babar~\cite{BabarXrad} and Belle~\cite{BelleXomega}. This observation
implies that the $X(3872)$ has positive C-parity. \babar\ reports an
evidence of the decay into $\psi(2S)\gamma$ final
state~\cite{BabarXrad} at $3.5\sigma$ level , using a
data sample of $424~\invfb$. The measured branching fraction ratio
$\calB(X(3872)\to\psi(2S)\gamma)/ \calB(X(3872)\to\jpsi\gamma) = 3.4 
\pm 1.4$, where the error is the sum in quadrature of the
statistical and systematic uncertainties, is
unexpectedly  large.
Recently, Belle reported a preliminary
result~\cite{belleXradPrel} of a search for $\psi(2S)\gamma$ decay
using $849~\invfb$ of data. No statistically significant signal is
observed. The branching fraction upper limit at 90\% confidence level
is $\calB(X(3872)\to\psi(2S)\gamma)/ \calB(X(3872)\to\jpsi\gamma) < 2.1$.

A common agreement concerning the interpretation of $X(3872)$
has not been reached. Possible conventional charmonium assignments are
$\chi_{c1}^{\prime}$ or $\eta_{c2}(1D)$. The first is challenged
by the fact that $\chi_{c2}^{\prime}$ is tentatively identified as
the $Z(3930)$. This implies that the mass splitting in the
$\chi_c^{\prime}$ triplet should be larger
than expected. However, it was noticed that the effect of the coupling
to $D\overline{D}^*$ channel may shift the $1^{++}$ state mass down to
$D\overline{D}^{*}$ threshold, while the shift of the $2^{++}$ state
is expected to be much smaller~\cite{xcc}. 
The $\eta_{c2}(1D)$ interpretation is favored by \babar\
assignment of $J^{PC}=2^{-+}$. The measured $X(3872)$ mass is
consistent with the $1^1D_2$ state mass predicted
in~\cite{etacdmass}, but not consistent with the values reported in
many recent papers~\cite{etacdnomass}. 
Theoretical calculations~\cite{etacdrad,etacdfer} shows that the predicted
decay rate of such a state to $\psi(2S)\gamma$ is inconsistent with
the results reported by \babar~\cite{BabarXrad}. It should be noted
that recent Belle preliminary result for this decay mode disagrees with
\babar\ one~\cite{belleXradPrel}.  Recently it has been noticed
that also the decay rate to $D^0\overline{D}^{*0}$ cannot fit the $1^1D_2$
assignment~\cite{etacdfer}. 
$X(3872)$ interpretation as a $D^0\overline{D}^{*0}$ molecule was
suggested by many 
authors~\cite{page,swanson,Xmolecule}, and may accommodate the observed 
large isospin violation~\cite{swanson}. However, the decays into
$\jpsi\gamma$ and 
$\psi(2S)\gamma$ imply that this molecular state should mix with
$\chi_{c1}^{\prime}$~\cite{swansonMix}. Such a mixing may be not as
large as proposed, in the light of the new Belle result on
$\psi(2S)\gamma$ decay~\cite{belleXradPrel}. Molecular interpretation is 
challenged by \babar\ $J^{PC}=2^{-+}$ assignment. Furthermore, it was
noticed that the molecular interpretation  may be
inconsistent with the large cross section for prompt production that
can be derived from CDF results~\cite{polosaCDF}. 
The interpretation of $X(3872)$ as a tetraquark~\cite{maiani} predicts
the presence of neutral and charged partners in the same mass
region. Such states were searched and not found~\cite{Xtetra}.

\section{The 3940 Family}
The $Y(3940)$ was observed by Belle in the
$B\to\jpsi\omega K$ process~\cite{Belle3940}, and confirmed by
\babar~\cite{Babar3940}. The same structure is not observed in the $B\to
D^0\overline{D}^{*0}K$ process~\cite{Xtoddst}. Belle measures the mass
$(3943\pm11(stat)\pm13(syst))$~\mevcc\ 
and width $(87\pm22(stat)\pm26(syst))$~\mevcc.
A resonance in the the $\jpsi\omega$ final state, dubbed
$X(3915)$, was observed by Belle in two-photon
collisions~\cite{Belle3915}. The 
measured mass and width are $(3914\pm3(stat)\pm2(syst))$~\mevcc\ and
$(17\pm10(stat)\pm3(syst))$~\mevcc, respectively. A recent  
\babar\ re-analysis~\cite{BabarXomega} provides a Y(3940) mass and
width of $(3919.1^{+3.8}_{-3.5}(stat)\pm2.0(syst))$~\mevcc\ and
$(31^{+10}_{-8}(stat)\pm5(syst))$~\mevcc,  respectively. 
This latter measurement favors a $Y(3940)$ mass slightly lower 
than the one first reported by Belle~\cite{Belle3940}. This value is
in agreement with the one measured by Belle for the $X(3915)$ in the
two-photon process. Thus, it seems 
likely that the same particle, with a mass of about 3915~\mevcc, is
observed in two distinct production processes. Belle reports a product
of the two-photon width times the decay branching ratio
$\Gamma_{\gamma\gamma}(Y(3940))\times\calB(Y(3940)\to\jpsi\omega)$ equal to
$(61\pm17(stat)\pm8(syst))$~eV and $(18\pm5(stat)\pm2(syst))$~eV 
for $J^{P}=0^+$  and $2^+$ assignments,
respectively~\cite{Belle3915}. Assuming 
$\Gamma_{\gamma\gamma}(Y(3940))\sim1~\mbox{keV}$, that is a typical
value for excited charmonium, the branching ratio
$\calB(Y(3940)\to\jpsi\omega)$ is in the range 1--6\%, which is
unexpectedly large, compared to other excited \ccbar\
states~\cite{PDG}. The proposed interpretation of $Y(3940)$ as the
$\chi_{c1}^{\prime}$ state, where the final state interaction enhances
the $\jpsi\omega$ decay~\cite{eichten}, is ruled out by the
observation of this state in two-photon production. Interpretation as
$\chi_{c0}^{\prime}$ was also suggested~\cite{newPwaves}.
Interpretation as a charmonium hybrid is seriously challenged by 
lattice calculations that show that the expected mass for hybrid
ground state should be some 500~\mevcc\ higher than the one of
$Y(3940)$.
Interpretation in the framework
of molecular model has been proposed~\cite{Ymolec,Branz,YmolecGamma}. It
was suggested that the decay into $D^{*}\overline{D}\gamma$ may give
more insight on the nature of this state~\cite{YmolecGamma}.

Another resonant state, dubbed $Z(3930)$, was observed by 
Belle~\cite{BelleZ} and confirmed by 
\babar~\cite{BabarZ}. It was studied in 
two-photon production process and $D\overline{D}$ final
state. The measured mass and width are
$(3929\pm5(stat)\pm2(syst))~\mevcc$ and 
$(29\pm10(stat)\pm2(syst))~\mevcc$ for Belle and
$(3927\pm2(stat)\pm1(syst))~\mevcc$  
and $(21\pm7(stat)\pm4(syst))~\mevcc$ for \babar. The agreement between
the two measurements is very 
good. The decay into spinless final state, combined with the fact that
$C=+1$ due to the production mechanism, implies that $J=L$ is even and
thus $P=+1$. The value of $J$ is determined by studying the angular
distribution of the decay products. In particular, the angle between
the directions of the $D\overline{D}$ system and the beam provides
such kind of information. Both experiments favor $J^{PC}=2^{++}$
assignment. 
Belle provides a measurement of the branching fraction ratio
$\calB(Z(3930)\to D^+D^-)/\calB(Z(3930)\to D^0\overline{D}^0) = 0.74 \pm
0.43 (stat)\pm 0.16(syst)$ which suggests isospin invariance, as
expected for conventional \ccbar.
The product of the two-photon width times the decay branching ratio
$\Gamma_{\gamma\gamma}(Z(3930))\times\calB(Z(3930)\to D\overline{D})$ is
found to be $(0.18\pm0.06)$~keV and $(0.24\pm0.05)$~keV, by Belle and
\babar, respectively, where the error is the sum in quadrature of the
statistical and systematic uncertainties. 
There is a general agreement about the interpretation
of this state, which is identified as $\chi_{c2}^{\prime}$. The mass,
two-photon width and decay angular distribution are consistent with
theoretical expectation for this charmonium
state~\cite{etacdmass,3930theor}. 

\section{New $\jpsi\phi$ States}
The search of structures in the $\jpsi\phi$ mass spectrum is motivated
by the prediction that a $c\bar{c}s\bar{s}$ tetraquark is expected to
have sizable branching ratio in this final state and a mass in the
4270--4350~\mevcc\ range~\cite{ccssMass}.
CDF has reported an evidence of a resonance-like candidate, dubbed
$Y(4140)$, with a significance of $3.8\sigma$~\cite{CDFPsiPhi}.  
The $\jpsi\phi$ final state was studied in $\Bp\to\jpsi\phi K^+$ decay.
The measured mass and width are
$(4143.0\pm2.9(stat)\pm1.2(syst))~\mevcc$ and
$(11.7^{+8.3}_{-5.0}(stat)\pm3.7(syst))~\mevcc$. A subsequent document by 
CDF estimates the branching fraction product $\calB(\Bp\to
Y(4140)K^+)\times\calB(Y(4140)\to\jpsi\phi)=(9.0\pm3.4(stat)\pm2.9(syst))\times 
10^{-6}$~\cite{CDF4140bf}. Belle has searched for the $Y(4140)$ using
the same production mechanism, and found no evidence of
it~\cite{BellePsiPhi}. However, due to small detection efficiency near
the $\jpsi\phi$ threshold, the upper limit on the branching ratio
$\calB(\Bp\to
Y(4140)K^+)\times\calB(Y(4140)\to\jpsi\phi)<6\times10^{-6}$ is not in
contradiction with CDF measurement, considering its large
uncertainty. Several interpretation were proposed for $Y(4140)$,
including a $D^{*+}_sD^{*-}_s$
molecule~\cite{Branz,4140molecule,huang,4140hybrid}, an exotic $1^{-+}$ 
hybrid~\cite{4140hybrid},  a $c\bar{c}s\bar{s}$
tetraquark~\cite{4140tetra}, and an effect of the $\jpsi\phi$
threshold opening~\cite{4140threshold}. Some arguments were raised
against the interpretation as a standard \ccbar\ state~\cite{4140notcc}
and scalar  $D^{*+}_sD^{*-}_s$ molecule~\cite{4140notmolecule}. 
Authors of Ref.~\cite{Branz} predict, in the $D^{*+}_sD^{*-}_s$ molecule
picture, the product of the two-photon width times the decay branching ratio
$\Gamma_{\gamma\gamma}(Y(4140))\times\calB(Y(4140)\to\jpsi\phi)$ to be
sizable, with large theoretical uncertainties. Belle searched for the
$Y(4140)$ in two-photon production and found no evidence of
it~\cite{BellePsiPhigg}. Furthermore, Belle has reported evidence of  a
narrow structure with a mass equal to
$(4350^{+4.6}_{-5.1}(stat)\pm0.7(syst))~\mevcc$ and width
$(13^{+18}_{-9}(stat)\pm4(syst))~\mevcc$. The structure, dubbed
$X(4350)$, has a significance of $3.2\sigma$. The measured mass is
inconsistent with the $Y(4140)$ one. 
Interpretation of the $X(4350)$ as $\chi_{c2}^{\prime\prime}$ was
suggested~\cite{newPwaves}. Other interpretations as an exotic state
are similar to the $Y(4140)$ ones~\cite{huang,4140tetra,4350int}.  

\section{Charged States}
The observation of a resonance-like state with non-zero net electric
charge would be striking evidence of a state with an unconventional
nature, 
since it would be inconsistent with the \qqbar\ structure. In
particular, both multiquark states and \emph{hadrocharmonium} may
accommodate such non-zero charge. Belle reported the evidence of the
narrow $Z(4430)^-$ decaying into $\psi(2S)\pip$, with a
significance of 5.4$\sigma$~\cite{BelleZchDP}. This state was 
studied in the decay $\Bz\to\psi(2S)\pim K^+$. The Dalitz plot of this
decay is dominated by the presence of the $K^{*}$ resonances. In
Belle's first analysis, a veto was applied to remove such
contributions~\cite{BelleZch}. In a more recent analysis, a full Dalitz
plot analysis 
was performed~\cite{BelleZchDP}. Both analyses report the same value
for the mass
$(4433^{+15}_{-12}(stat)^{+19}_{-13}(syst))$~\mevcc~\cite{BelleZchDP}.     
The measured widths are equal to 
$(45^{+18}_{-13}(stat)^{+30}_{-13}(syst))$~\mevcc\ and 
$(107^{+86}_{-43}(stat)^{+74}_{-56}(syst))$~\mevcc, in the
first~\cite{BelleZch} and latter~\cite{BelleZchDP} analysis,
respectively. These 
measurements are consistent inside the large uncertainties of the 
latter. \babar\ has searched for the $Z(4430)$ in both $\jpsi\pim$ and
$\psi(2S)\pim$ final states, but no evidence of resonance-like
structures has been found~\cite{BabarZch}. In this analysis, \babar\
has performed a study of the reflections of the $K^*$ system in the
$\jpsi(\psi(2S))\pim$ mass spectrum, and found that such reflections
reproduce data well, without the need of any additional resonant
structures.  
The upper limit at 90\% confidence level on the branching
fraction product
$\calB(\Bz\to Z(4430)^-K^+)\times\calB(Z(4430)^-\to\psi(2S)\pim)$
reported by \babar\ is $3.1\times10^{-5}$. This is not in contrast with
the measured value
$(3.2^{+1.8}_{-0.9}\phantom{}^{+5.3}_{-1.6})\times10^{-5}$ reported by
Belle.    
No analysis of the \jpsi\pim\ final state has been reported by 
Belle so far.  

Belle reported evidence of two more states ($Z_1(4050)^-$ and
$Z_2(4250)^-$) with non-zero electric charge in the final state
$\chi_{c1}\pim$~\cite{BelleZchich}. These states were found in 
$B\to\chi_{c1}\pim K$, whose Dalitz plot is dominated by the $K^*$
resonances. A Dalitz plot  analysis is performed. The solution with
two resonant structures is favored  with respect to the one with no
resonant contributions, with a 
significance of $5.7\sigma$. The mass of these 
states is $(4051\pm14(stat)^{+20}_{-41}(syst))$~\mevcc\ and
$(4248^{+44}_{-28}(stat)^{+180}_{-35}(syst))$~\mevcc, 
respectively. Their width is
$(82^{+21}_{-17}(stat)^{+47}_{-22}(syst))$~\mevcc\ and
$(177^{+54}_{-39}(stat)^{+316}_{-61}(syst))$~\mevcc, respectively.   
No search for such states has been reported by \babar\ so far.

\section{Conclusions}
In conclusion, recent advances on both theoretical and experimental
sides concerning charmonium spectroscopy have brought renewed interest in
this field. Unusual properties of newly discovered states may suggest
that they represent some form of hadronic matter predicted by QCD but
never observed so far, such as multiquarks or molecular states.
Some of the new states are well established 
experimentally, such as $X(3872)$ and $Y(4260)$, but the
determination of their properties and the theoretical interpretation
is still an open issue. In particular, the recent $X(3872)$
measurement in the $\jpsi\omega$ final state by
\babar~\cite{BabarXomega} favors a
$J^{PC}=2^{-+}$  assignment, opposed to $1^{++}$ as
generally accepted previously.    
New measurements of the properties of such a state, also at hadronic
machines, may shed light on its nature.
Many other states have been claimed, but are missing confirmation. Among
them, the most interesting are the $Z^-$ states that exhibit non-zero
electric charge. A confirmation of such states would be a compelling
evidence of the existence of mesons with non-\qqbar\ structure. The
observation of $Z(4430)$ was claimed by Belle~\cite{BelleZch} and not
confirmed by \babar~\cite{BabarZch}. This state is in principle
accessible at the Tevatron and LHC 
experiments. Alternatively, the search for $Z_1(4050)^-$ and
$Z_2(4250)^-$ at hadronic machines could be quite tricky, or even
unfeasible, since the $\chi_{c1}$ is reconstructed  through its
$\chi_{c1}\to\jpsi\gamma$ decay, where the photon has an energy of
about $410$~\mev. An ideal environment to perform measurements of exotic
charmonia properties and searches for new states would be the future
super-flavor factories SuperB~\cite{superb} and
SuperKEKB~\cite{superkek}. 

\section{Acknowledgements}
I would like to thank my \babar\ collaborators and in particular
F.~Palombo, B.~Fulsom, and S.~Stracka for their help in improving my
talk and this proceeding contribution.


\begin{thebibliography}{99}
\bibitem{jpsi}
J.~J.~Aubert \etal\ \jprl{33}, 1404 (1974); 
J.-E.~Augustin \etal\ \jprl{33}, 1406 (1974). 

\bibitem{qwg}
Quarkonium Working Group, N.~Brambilla \etal\ [hep-ph/0412158].

\bibitem{BelleX}
Belle Collaboration, S.-K.~Choi \etal\ \jprl{91}, 262001 (2003)
[hep-ex/0309032].
 
\bibitem{CDFX}
CDF Collaboration, D.~E.~Acosta \etal\ \jprl{93}, 072001 (2004)
[hep-ex/0312021].

\bibitem{D0X}
D0 Collaboration, V.~M.~Abazov \etal\ \jprl{93}, 162002 (2004)
[hep-ex/0405004].

\bibitem{BabarX}
\babar\ Collaboration, B.~Aubert \etal\ \jprd{71}, 071103 (2005)
[hep-ex/0406022].

\bibitem{BabarXwidth}
\babar\ Collaboration, B.~Aubert \etal\  \jprd{77}, 011102(R) (2008)
[hep-ex/0708.1565].

\bibitem{BabarY}
\babar\ Collaboration, B.~Aubert \etal\  \jprl{95}, 142001 (2005)
[hep-ex/0506081].

\bibitem{CLEOY}
CLEO Collaboration, T.~E.~Coan \etal\ \jprl{96}, 162003 (2006)
[hep-ex/0707.2541].

\bibitem{BelleY}
Belle Collaboration, C.~Z.~Yuan \etal\ \jprl{99}, 182004 (2007)
[hep-ex/0707.2541].

\bibitem{OtherY}
\babar\ Collaboration, B.~Aubert \etal\ \jprl{98}, 212001 (2007)
[hep-ex/0610057]; \\
Belle Collaboration, X.~L.~Wang \etal\  \jprl{99}, 142002 (2007)
[hep-ex/0707.3699].

\bibitem{Belle3915}
Belle Collaboration, S.~Uehara \etal\  \jprl{104}, 092001 (2010)
[hep-ex/0912.4451].

\bibitem{BelleDC}
Belle Collaboration, K.~Abe \etal\ , \jprl{98}, 082001 (2007)
[hep-ex/0507019];\\
Belle Collaboration, P.~Pakhlov \etal\  \jprl{100}, 202001 (2008)
[hep-ex/0708.3812].

\bibitem{Belle3940}
Belle Collaboration, S.-K.~Choi \etal\   \jprl{94}, 182002 (2005)
[hep-ex/0408126].

\bibitem{Babar3940}
\babar\ Collaboration, B.~Aubert \etal\ \jprl{101}, 082001 (2008)
[hep-ex/0711.2047].

\bibitem{BelleZ}
Belle Collaboration, S.~Uehara \etal\  \jprl{96}, 082003 (2006)
[hep-ex/0512035].

\bibitem{BabarZ}  
\babar\ Collaboration, B.~Aubert \etal\ \jprd{81}, 092003 (2010)
[hep-ex/1002.0281].

\bibitem{CDFPsiPhi}
CDF Collaboration, T.~Aaltonen \etal\ \jprl{102}, 242002 (2009)
[hep-ex/0903.2229].

\bibitem{BellePsiPhi}
C.-Z.~Yuan (on behalf of the Belle Collaboration), \emph{XXIX Physics in
Collision}, Proceedings of the International Symposium in Kobe, Japan,
August 30 - September 2, 2009, [hep-ex/0910.3138].

\bibitem{BellePsiPhigg}
Belle Collaboration, C.~P.~Shen \etal\  \jprl{104}, 112004 (2010)
[hep-ex/0912.2383].

\bibitem{BelleZch}
Belle Collaboration, S.-K.~Choi \etal\  \jprl{100}, 142001 (2008)
[hep-ex/0708.1790].

\bibitem{BabarZch}
\babar\ Collaboration, B.~Aubert \etal\  \jprd{79}, 112001 (2009)
[hep-ex:/0811.0564].

\bibitem{BelleZchich}
Belle Collaboration, R.~Mizuk \etal\ \jprd{78}, 072004 (2008)
[hep-ex/0806.4098].

\bibitem{hybrids}
P.~Hasenfratz \etal\ \plb{95}, 299 (1980); 
S.~Perantonis and C.~Michael, \emph{Nucl. Phys. B} {\bf 347}, 854
(1990); 
S.~Ishida \etal\ \jprd{47}, 179 (1993) and references
therein; 
T.~Barnes \etal\ \jprd{52} 5242 (1995)[hep-ph/9501405];
F.~E.~Close and P.~R.~Page,  \emph{Nucl. Phys. B} {\bf 443}, 233
(1995)[hep-ph/9411301];
Y.~S.~Kalashnikova and D.~S.~Kuzmenko, \emph{Phys. Atom. Nucl.}{\bf
  66}, 955 (2003)[hep-ph/0203128].

\bibitem{polosa}
N.~Drenska \etal\ [hep-ph/1006.2741] and references threin;

\bibitem{tetraquarks}
D.~Ebert \etal\ \plb{634}, 214 (2006)[hep-ph/0512230]; \jprd{76}, 114015
(2007)[hep-ph/0706.3853]; 
T.~W~Chiu and T.~H.~Hsieh,  \plb{646}, 95 (2007)[hep-ph/0603207] and
references therein;
L.~Liu, PoS(LAT2009), 99;
Z.-G~Wang \etal\ [hep-ph/1004.0484] and referneces therein.

\bibitem{pentaquarks}
R.~Jaffe and F.~Wilczek, \jprl{91}, 232003 (2003)[hep-ph/0307341];
M.~Abud \etal\ \emph{Adv. Stud. Theor. Phys.}, {\bf 2}, 929
(2008)[hep-ph/0806.4581]. 

\bibitem{exa}
M.~Abud \etal\ \jprd{81}, 074018 (2010)[hep-ph/0912.4299].

\bibitem{molecules}
N.~A.~Tornqvist, \jprl{67}, 556 (1991); \emph{Z. Phys. C}, {\bf 61},
525 (1994); \plb{590} 209 (2004)[hep-ph/0402237];
E.~Braaten and M.~Kusunoki, \jprd{69}, 074005 (2004)[hep-ph/0311147]; 
\emph{Phys. Rep.}, {\bf 429}, 243 (2006)[hep-ph/0601110];
F.~Close and C.~E.~Thomas, \jprd{78}, 034007 (2008)[hep-ph/0805.3653];
F.~Close \etal\ \jprd{81}, 074033 (2010)[hep-ph/1001.2553].

\bibitem{page}
F.~Close and P.~R.~Page, \plb{578}, 199 (2004)[hep-ph/0309253].

\bibitem{swanson}
E.~S.~Swanson, \plb{588}, 189 (2004)[hep-ph/0311229].

\bibitem{hadrocharmonium}
S.~Dubynskiy and M.~B.~Voloshin, \plb{666}, 344
(2008)[hep-ph/0803.2224]. 

\bibitem{threshold}
D.~V.~Bugg, \jpg{35}, 075005 (2008)[hep-ph/0802.0934];
\emph{Int. J. Mod. Phys.A} {\bf 24}, 394 (2009) and references
therein. 

\bibitem{eichten}
E.~Eichten \etal\ \emph{Rev. Mod. Phys.} {\bf 80}, 1161
(2008)[hep-ph/0701208].

\bibitem{babarNIM}
\babar\ Collaboration, B.\ Aubert \etal\ \nima{479}, 1 (2002)
[hep-ex/0105044]. 

\bibitem{belleNIM}
Belle Collaboration, \nima{479}, 117 (2002) 

\bibitem{cdfNIM}
CDF Collaboration, F.~Abe \etal\ \nima{271}, 387 (1988);
CDF Collaboration, R.~Blair \etal\ Fermilab Report
No. FERMILAB-Pub-96/390-E (1996).

\bibitem{yang}
C.~N.~Yang, \pr{77}, 242 (1950).

\bibitem{CDFrho}
CDF Collaboration, A.~Abulencia \etal\ \jprl{96}, 102002 (2006)
[hep-ex/0512074].

\bibitem{CDFang}
CDF Collaboration, A.~Abulencia \etal\ \jprl{98}, 132002
(2007)[hep-ex/0612053]. 

\bibitem{BelleXang}
Belle Collaboration, K.~Abe \etal\ [hep-ex/0505038] (2005).

\bibitem{rosner}
J.~L.~Rosner, \jprd{70}, 094023 (2004)[hep-ph/0408334];
D.~V.~Bugg, \jprd{71}, 016006 (2005)[hep-ph/0410168].

\bibitem{BelleXddp}
Belle Collaboration, G.~Gokhroo \etal\ \jprl{97}, 162002
(2006)[hep-ex/0606055]; 

\bibitem{Xtoddst}
\babar\ Collaboration, B.~Aubert \etal\ \jprd{77}, 011102(R)
(2008)[hep-ex/0708.1565]; \\
Belle Collaboration, T.~Aushev \etal\ \jprd{81}, 031103 (2010)
[hep-ex/0810.0358].


\bibitem{BelleXomega}
Belle Collaboration, K.~Abe \etal\ [hep-ex/0505037] (2005).

\bibitem{BabarXomega}
\babar\ Colalboration, P.~del~Amo~Sanchez \etal\ \jprd{82}, 011101(R)
(2010)[hep-ex/1005.5190]. 

\bibitem{BabarXrad}
\babar\ Collaboration, B.~Aubert \etal\ \jprl{102}, 132001
(2009)[hep-ex/0809.0042]. 

\bibitem{belleXradPrel}
V.~Bhardwaj (on behalf of the Belle Collaboration), talk at
\emph{International Workshop on Heavy Quarkonium (QWG)}, 
FERMILAB, Batavia, IL, May 18-21, 2010.

\bibitem{xcc}
Yu.~S.~Kalashnikova, \jprd{72}, 034010 (2005)[hep-ph/0506270];
Yu.~S.~Kalashnikova and A.~V.~Nefediev, \jprd{80}, 074004
(2009)[hep-ph/0907.4901];  
I.~V.~Danilkin and Yu.~A.~Simonov, \jprd{81}, 074027
(2010)[hep-ph/0907.1088]; [hep-ph/1006.0211] (2010);
S.Coito \etal\ [hep-ph/1008.5100] (2010).

\bibitem{etacdmass}
S.~Godfrey and N.~Isgur, \jprd{32}, 189 (1985).

\bibitem{etacdnomass}
A.~M.~Badalian \etal\ Phys. At. Nucl. {\bf 63}, 1635 (200);
T.~Barnes \etal\ \jprd{72}, 054026 (2005);
T.~J.~Burns \etal\ [hep-ph/1008.0018] (2010).

\bibitem{etacdrad}
Y.~Jia \etal\ [hep-ph/1007.4541] (2010).

\bibitem{etacdfer}
Yu.~S.~Kalashinkova and A.~V.~Nefediev, [hep-ph/1008.2895] (2010).

\bibitem{Xmolecule}
M.~B.~Voloshin, \plb{579}, 361 (2004)[hep-ph/0309307];
E.~Braaten and M.~Kusunoki, \jprd{72}, 054022 (2005)[hep-ph/0507163].

\bibitem{swansonMix}
E.~S.~Swanson, \plb{598}, 197 (2004)[hep-ph/0406080].

\bibitem{polosaCDF}
C.~Bignamini \etal\ \jprl{103}, 162001 (2009)[hep-ph/0906.0882]. 

\bibitem{maiani}
L.~Maiani \etal\ \jprd{71}, 014028 (2005)[hep-ph/0412098].

\bibitem{Xtetra}
\babar\ Collaboration, B.~Aubert \etal\ \jprd{71}, 031501(R)
(2005)[hep-ex/0412051];  \\
CDF Collaboration, A.~Aaltonen \etal\ \jprl{103}, 152001
(2009)[hep-ex/0906.5218]. 

\bibitem{PDG}
C.~Amsler \emph{et al.} (Particle Data Group), \plb{667}, 1 (2008),   
and 2009 partial update for the 2010 edition. 

\bibitem{newPwaves}
X.~Liu \etal\ \jprl{104}, 122001 (2010)[hep-ph/0911.3694].

\bibitem{Ymolec}
X. Liu \etal\ \epjc{61}, 411 (2009)[hep-ph/0808.0073].

\bibitem{Branz}
T Branz \etal\ \jprd{80}, 054019 (2009)[hep-ph/1001.3959].

\bibitem{YmolecGamma}
W.~H.~Liang \etal\ \epja{44}, 479 (2010)[hep-ph/0912.4359].

\bibitem{3930theor}
C.~R.~Munz, \emph{Nucl. Phys. A} {\bf 609}, 364
(1996)[hep-ph/9601206];  
E.~Eichten \etal\ \jprd{69}, 094019 (2004)[hep-ph/0401210]. 

\bibitem{4140molecule}
X.~Liu and S.~L.~Zhu, \jprd{80}, 017502 (2009)[hep-ph/0903.2529];
G.~J.~Ding, \epjc{64} 297, (2009)[hep-ph/0904.1782];
X.~Liu and H.~W.~Ke, \jprd{80}, 034009 (2009)[hep-ph/0907.1349];
R.~M.~Albuquerque \etal\ \plb{678}, 186 (2009)[hep-ph/0903.5540];
R.~Molina and E.~Oset, \jprd{80}, 114013 (2009)[hep-ph/0907.3043].

\bibitem{huang}
J.~R.~Zhang and M.~Q.~Huang, \jprd{80}, 056004
(2009)[hep-ph/0906.0090]; \emph{J. Phys. G} {\bf 37}, 025005 (2010)
[hep-ph/0905.4178]; 

\bibitem{4140hybrid}
N.~Mahajan, \plb{679}, 228 (2009)[hep-ph/0903.3107].
 
\bibitem{4140tetra}
Fl.~Stancu, \emph{J. Phys. G} {\bf 37}, 075017
(2010)[hep-ph/0906.2485]. 

\bibitem{4140threshold}
E.~van~Beveren and G.~Rupp [hep-ph/0906.2278] (2009).

\bibitem{4140notcc}
X.~Liu, \plb{680}, 137 (2009)[hep-ph/0904.0136].

\bibitem{4140notmolecule}
Z.-G.~Wang, \epjc{63}, 115 (2009)[hep-ph/0903.5200];
Z.-G.~Wang \etal\ \epjc{64}, 373 (2009)[hep-ph/0907.1467].

\bibitem{4350int}
Y.-L.~Ma [hep-ph/1006.1276] (2010);
A.~Albuquerque \etal\ \plb{690} 141 (2010)[hep-ph/1001.3092];
Z.-G.~Wang, \plb{690}, 403 (2010)[hep-ph/0912.4626].

\bibitem{ccssMass}
N.~V.~Drenska \etal\ \jprd{79}, 077502 (2009)[hep-ph/0902.2803]

\bibitem{CDF4140bf}
K.~Yi (on behalf of the CDF Collaboration), proceedings of DPF-2009,
Detroit, MI, July 2009, eConf~C090726 [hep-ex/0910.3163]. 

\bibitem{BelleZchDP}
Belle Collaboration, R.~Mizuk \etal\ \jprd{80}, 031104(R)
(2009)[hep-ex/0905.2869].

\bibitem{superb}
SuperB Collaboration, M.~Bona \etal\ [hep-ex/0709.0451] (2007).
SuperB Collaboration, B.~O'Leary \etal\ [hep-ex/1008.1541] (2010).

\bibitem{superkek}
A.~G.~Akeroyd \etal\ [hep-ex/1002.5012] (2010).

\end{thebibliography}
\end{document}